\documentclass[usenatbib,fleqn,letters]{mnras}
\usepackage{showyourwork}
\usepackage{graphicx}
\usepackage{amssymb,amsmath}
\usepackage{color}
\usepackage{multirow}
\usepackage{soul}
\usepackage{orcidlink}
\usepackage{comment}

\newcommand{\sarc}{$^{\prime\prime}\!\!$}

\newcommand{\wphz}{$\,$W$\,$Hz$^{-1}$}

\newcommand{\muJybm}{$\mu$Jy$\,$beam$^{-1}$}
\newcommand{\llof}{$L_{\textrm{144MHz}}$}

\defcitealias{morabito_identifying_2022}{M22}
\defcitealias{best_lofar_2023}{B23}

\title[A hidden AGN population]{A hidden Active Galactic Nuclei population: the first radio luminosity functions constructed by physical process}

\author[L.K. Morabito]{\parbox{\textwidth}{Leah K. Morabito$^{1,2}$\thanks{E-mail: leah.k.morabito@durham.ac.uk}\orcidlink{0000-0003-0487-6651}
R. Kondapally$^{3}$\orcidlink{0000-0001-6127-8151},
P.N. Best$^{3}$\orcidlink{0000-0001-5081-4801},
B.-H. Yue$^{3}$\orcidlink{0009-0009-8935-2929}, 
J.M.G.H.J. de Jong$^{4}$\orcidlink{0000-0001-6876-8719},
F. Sweijen$^{1}$\orcidlink{0000-0002-6470-7967},
Marco Bondi$^{5}$\orcidlink{0000-0002-9553-7999},
Dominik J. Schwarz$^{6}$\orcidlink{0000-0003-2413-0881},
D.J.B. Smith$^{7}$\orcidlink{0000-0001-9708-253X},
R.J. van Weeren$^{4}$\orcidlink{0000-0002-0587-1660},
H.J.A. R\"{o}ttgering$^{4}$\orcidlink{0000-0001-8887-2257},
T.W. Shimwell$^{4,8}$\orcidlink{0000-0001-5648-9069},
Isabella Prandoni$^{5}$\orcidlink{0000-0001-9680-7092}
\\}\\ 
$^{1}$Centre for Extragalactic Astronomy, Department of Physics, Durham University, Durham DH1 3LE, UK \\
$^{2}$Institute for Computational Cosmology, Department of Physics, Durham University, South Road, Durham DH1 3LE, UK \\ 
$^{3}$Institute for Astronomy, University of Edinburgh Royal Observatory, Blackford Hill, Edinburgh, EH9 3HJ, UK \\
$^{4}$Leiden Observatory, Leiden University, PO Box 9513, 2300 RA Leiden, The Netherlands \\
$^{5}$INAF - Istituto di Radioastronomia, Via Gobetti 101, 40129 Bologna, Italy  \\
$^{6}$Fakult\"{a}t f\"{u}r Physik, Universit\"{a}t Bielefeld, Postfach 100131, 33501 Bielefeld, Germany \\
$^{7}$Centre for Astrophysics Research, University of Hertfordshire, College Lane, Hatfield AL10 9AB, UK \\
$^{8}$ASTRON, Netherlands Institute for Radio Astronomy, Oude Hoogeveensedijk 4, 7991 PD, Dwingeloo, the Netherlands \\}

\begin{document}

\date{}
\pagerange{\pageref{firstpage}--\pageref{lastpage}} \pubyear{2024}
\maketitle

\label{firstpage}

\begin{abstract}
Both star formation (SF) and Active Galactic Nuclei (AGN) play an important role in galaxy evolution. Statistically quantifying their relative importance can be done using radio luminosity functions. Until now these relied on galaxy classifications, where sources with a mixture of radio emission from SF and AGN are labelled as either a star-forming galaxy or an AGN. This can cause the misestimation of the relevance of AGN. Brightness temperature measurements at 144 MHz with the International LOFAR telescope can separate radio emission from AGN and SF. We use the combination of sub-arcsec and arcsec resolution imaging of 7,497 sources in the Lockman Hole and ELAIS-N1 fields to identify AGN components in the sub-arcsec resolution images and subtract them from the total flux density, leaving flux density from SF only. We construct, for the first time, radio luminosity functions by physical process, either SF or AGN activity, revealing a hidden AGN population at \llof$<10^{24}$\wphz . This population is 1.56$\pm$0.06 more than expected for $0.5<z<2.0$ when comparing to RLFs by galaxy classification. The star forming population has only 0.90$\pm$0.02 of the expected SF. These ‘hidden’ AGN can have significant implications for the cosmic star formation rate and kinetic luminosity densities.
\end{abstract}

\begin{keywords}
galaxies: active -- galaxies: star formation -- galaxies: evolution -- radio continuum: galaxies 
\vspace{3.7in}
\end{keywords}

\section{Introduction}
\label{sec:intro}
The extent to which active galactic nuclei (AGN) impact galaxy formation is a major open question in astrophysics. It is clear that there is an influence, from both observational and theoretical standpoints. Observations have revealed tight scaling relations between the mass of a super-massive black hole and its host galaxy \citep[see, e.g.][and references therein]{kormendy_coevolution_2013}, while cosmological simulations require some form of AGN feedback \citep{bower_breaking_2006,croton_many_2006} to be able to reproduce the observed galaxy population. 
The cosmic histories of AGN activity and star formation (SF), and by extension the interplay between them, are interlinked. It is necessary to quantify the contribution of each process to understand the importance of either or both of them. Relying on overall galaxy classification can over- or under-estimate the contributions of each process; for example, by considering all of the radio emission in a galaxy classified as `star-forming' to be due to star formation, when some may be due to AGN. Quantifying these contributions is the goal of this letter. 

Decomposing AGN activity from star formation (SF) on a per-galaxy basis is a difficult problem at any wavelength. Efforts have been made to do this via mid-infrared (MIR) spectral decomposition using template fitting \citep[e.g.,][]{laurent_mid-infrared_2000,hernan-caballero_resolving_2015,li_active_2024}. Multi-wavelength approaches use spectral energy distribution (SED) fitting, which uses information from as many observed bands as possible to jointly fit galaxy and AGN components \citep[e.g.][]{calistro_rivera_agnfitter_2016,boquien_cigale_2019,pacifici_art_2023}. However, SED fitting is costly as it relies on observations from a variety instruments, which then have to be carefully cross-matched, and there can be degeneracy in the fits. Varying selection effects across the different bands can also present a problem. 

An alternative method to unambiguously identify the AGN component(s) in a galaxy is through brightness temperature, $T_b$, measurements in the radio waveband. Observations of radio emission have the advantage of being unobscured by dust or gas, so there are minimal observational biases. 
Models of radio emission from SF predict an upper limit to the surface brightness that can be produced, even in starburst galaxies \citep{condon_radio_1992}. Surface brightness which is detected above this limit therefore has to be attributed to AGN activity. Brightness temperature is inversely proportional to frequency, $\nu$, and resolution, $\theta$: $T_b \propto \nu^{-2}\theta^{-2}$. Traditionally brightness temperature measurements require Very Long Baseline Interferometry (VLBI) which reach milli-arcsec resolutions at GHz frequencies \citep[e.g.,][]{middelberg_mosaiced_2013,herrera_ruiz_faint_2017,radcliffe_nowhere_2018} to achieve enough $T_b$ sensitivity to make meaningful measurements. In \citet[][hereafter, \citetalias{morabito_identifying_2022}]{morabito_identifying_2022} we demonstrated that this can be done using sub-arcsecond resolution (0.\sarc\ 3) at 144 MHz with the LOw Frequency ARray \citep[LOFAR;][]{van_haarlem_lofar:_2013}. The advantage of LOFAR is its wide field of view: $\sim$6 deg$^2$ from a single observation versus a few to a couple hundred arcmin$^2$ with VLBI at GHz frequencies, depending on whether one images a single phase centre or uses multi-phase centres for correlation \citep{deller_difx-2:_2011,morgan_vlbi_2011}. For example, \citetalias{morabito_identifying_2022} used an 8 hour LOFAR observation of the Lockman Hole \citep{sweijen_deep_2022} to identify 940 AGN via brightness temperature, providing a sample two orders of magnitude larger than any VLBI sample of AGN identified at GHz frequencies, using a fraction of the observing time.

As part of the LOFAR Two-metre Sky Survey Deep Fields campaign \cite[][hereafter, \citetalias{best_lofar_2023}]{best_lofar_2023}, we have now doubled the areal coverage of 0.\sarc\ 3 resolution wide-field imaging. \cite{de_jong_into_2024} recently published their 0.\sarc\ 3 resolution image of the ELAIS-N1 field, which reaches a depth of 14 $\mu$Jy$\,$bm$^{-1}$ using 32 hours of observations. They also demonstrate the flexible resolution which can be achieved with LOFAR, by imaging at 0.\sarc\ 6 and 1.\sarc\ 2 resolution, which are both complementary to the existing 6\sarc\ \ Deep Fields Data Release 1 \citep[DR1;][]{sabater_lofar_2021,tasse_lofar_2021}. This is another advantage over VLBI instruments operating at GHz frequencies, which require either observing at different frequency bands or with other instruments to acquire information on the flux density of sources at other scales. As described in \S~5 of \citetalias{morabito_identifying_2022}, for sources which are unresolved in the 0.\sarc\ 3 resolution images and do not exhibit a radio excess over that predicted by the star formation rate (SFR) derived from SED fitting \citepalias[see][for more details]{best_lofar_2023}, we can separate the AGN activity from SF using a combination of the 0.\sarc\ 3 and 6\sarc\ \ resolution images.

The 2,483 sources in Lockman Hole \citep{sweijen_deep_2022} and now the 13,058 sources in ELAIS-N1 \citep{de_jong_into_2024} represent a massive step towards robust statistical studies where the population can be broken into bins of redshift and stellar mass to investigate AGN and SFG populations in detail. \cite{cochrane_lofar_2023} and \cite{kondapally_cosmic_2022} constructed radio luminosity functions (RLFs) from DR1 for the star forming galaxies and radio-excess AGN, respectively. Here we construct RLFs, for the first time, by physical process rather than overall galaxy classification. 
We investigate the cosmic evolution of AGN activity and SF by dividing the sample into redshift bins out to $z = 2.5$. 

In this paper, we first describe the data in \S~\ref{sec:data}, followed by an overview of the methods in \S~\ref{sec:methods}. Section~\ref{sec:results} describes the results, followed by conclusions in \S~\ref{sec:conclusions}. Throughout this paper, we use a $H_0 = 70$ km$\,$s$^{-1}\,$Mpc$^{-1}$, $\Omega_M = 0.3$, and $\Omega_{\Lambda}= 0.7$ cosmology. Radio flux density is defined as $S\propto \nu^{\alpha}$ where $\nu$ is frequency and $\alpha = -0.7$ is radio spectral index.

\section{Data}
\label{sec:data}

\subsection{LoTSS Deep Fields DR1}
\label{subsec:dr1}
The LOFAR Two-metre Sky Survey (LoTSS) has both a wide-area \citep{shimwell_lofar_2017,shimwell_lofar_2019,shimwell_lofar_2022} and Deep Fields \citep{tasse_lofar_2021,sabater_lofar_2021} component. The Deep Fields Data Release 1 (DR1) included Boötes, Lockman Hole, and ELAIS-N1, reaching sensitivities of 30, 23, and 20 \muJybm , using $\sim$80, $\sim$100, and $\sim$160 hours of observations, respectively. The resolution of these images is all 6\sarc . Alongside these deep radio images and their associated catalogues, \cite{kondapally_lofar_2021} provided a careful compilation of the ancillary data available at other wavebands (from far-infrared to ultraviolet) and cross-matching to the radio catalogues, followed by photometric redshifts in \cite{duncan_lofar_2021}. Collectively these catalogues provide an excellent resource, and we refer the reader for more details to the individual papers cited above.

Using the multi-wavelength cross-matched catalogues, \citetalias{best_lofar_2023} carried out detailed SED fitting for all radio-detected sources across the three deep fields. Full details can be found in their paper, but here we outline the most relevant details. All sources detected above 5$\sigma$ in their respective radio images had their far-infrared (FIR) to ultraviolet (UV) data fit using four different SED fitting software packages, and a `consensus' value was found for the derived galaxy properties using information from the goodness-of-fits.

The derived properties used in this paper are the consensus SFRs and the galaxy classifications which arise from the SED fitting. \citetalias{best_lofar_2023} use the empirical relationship between radio luminosity and SFR to define sources as either \textit{radio excess} or not. The AGN components used to fit the multi-wavelength information indicate whether a source is a radiatively efficient AGN (i.e., is there evidence for a disk/torus structure) or not. Combining these provides the following classes:
\begin{itemize}
    \item Radio excess + radiatively efficient (\textit{`radio-loud'})
    \item Radio excess + not radiatively efficient (\textit{`radio-loud'})
    \item Radiatively efficient + not radio excess (\textit{`radio-quiet'})
    \item Not radiatively efficient + not radio excess
\end{itemize}
The first three classes are AGN, and the final class is star forming galaxies (SFGs).  Five percent of all sources are unclassified, as they do not have a radio excess and the fits could not determine between radiatively efficient or inefficient. We note here that more recent work generated probabilistic classifications \citep{drake_lofar_2024} would have an impact on the RLFs, although the extent is a subject for future studies.

\subsection{High-resolution LOFAR imaging}
\label{subsec:highres}

Data for LoTSS are recorded with all available LOFAR stations, which are spread across eight different European countries, with a preponderance of stations in the Netherlands. The longest baseline (currently Ireland to Poland) is $\sim$2,000 km, providing unprecedented resolution at MHz frequencies. While LoTSS uses only the stations in the Netherlands for its standard data processing, post-processing of LoTSS data to achieve sub-arcsecond resolution is ongoing both for individual sources with $S>10\,$mJy in the wide-area survey \citep{morabito_sub-arcsecond_2022} and for the entire field of view in the Deep Fields \citep[][Escott et al., in prep; Bondi et al., in prep]{sweijen_deep_2022,de_jong_into_2024}. Due to the fact that the size of international stations is larger than that of stations in the Netherlands, the station beams have a smaller field of view. Coupled with bandwidth and time smearing, this limits the field of view effectively to a radius of 1.24 deg, which still provides 6.25 deg$^2$ of coverage (the imaged field of view is square with 2.5 deg per side). 

Both Lockman Hole and ELAIS-N1 have been imaged at 0.\sarc\ 3 resolution \citep[][respectively]{sweijen_deep_2022,de_jong_into_2024}. The resulting images have median rms noise of 34 \muJybm\ (Lockman Hole) and 17 \muJybm\ (ELAIS-N1), with the noise being lowest in the centre and increasing radially in both images. Although \cite{de_jong_into_2024} also imaged ELAIS-N1 at 0.\sarc\ 6 and 1.\sarc\ 2, for consistency with Lockman Hole we use only the 0.\sarc\ 3 resolution image. Catalogues from these images were processed to group individual Gaussian islands into complete sources, and remove duplicates. The resulting catalogues contain 2,316 and 13,058 sources above 5$\sigma$, for Lockman Hole and ELAIS-N1 respectively. 

Thanks to the added collecting area of the international stations, which increases the array's sensitivity, the 0.\sarc\ 3 resolution image for ELAIS-N1 has a lower noise level than the Deep Fields DR1. To ensure consistently built samples, and make use of DR1 information, we limit the high-resolution catalogues. For both fields, we first find the local rms in the companion 6\sarc\ \ resolution noise maps at the position of each source in the 0.\sarc\ 3 resolution catalogue. Using the peak brightness of the source in the catalogue from the 0.\sarc\ 3 resolution image, we calculate the peak-to-noise ratio and only select sources which would be detectable at the 5$\sigma$ limit in the 6\sarc\ resolution image, effectively ensuring all sources should have a match in DR1. We then cross-match based on radio position, to provide the radio information from the catalogues derived from both the 0.\sarc\ 3 resolution image and the 6\sarc\ \ resolution image. We use optical star masks to remove sources with multi-wavelength information affected by bright stars. Finally, we trim the area of the multi-wavelength catalogue to the 2.5 deg$\times$2.5 deg area which is covered by the 0.\sarc\ 3 resolution image (although the multi-wavelength coverage is smaller in ELAIS-N1 than in Lockman Hole). 
We are left with 2,252sources in Lockman Hole, and 5,245sources in ELAIS-N1. 

We determine whether sources are resolved or unresolved (relevant in Section~\ref{sec:methods}) in the same manner as \S~3.1 of \cite{shimwell_lofar_2019,shimwell_lofar_2022}, by finding an upper limit to the intrinsically unresolved sources. Theoretically, these should have equal peak brightness and total flux density, but LOFAR maps are impacted by smearing due to the ionosphere, and high resolution images have time and bandwidth smearing effects. Thus intrinsically unresolved sources can appear spatially extended. To find this upper limit, we take all sources which are fit by a single Guassian component and are smaller than a multiple of the beam size. For the Lockman Hole, we use a factor of 3, but for ELAIS-N1 we find that a factor of 4 provided a better sample. This may be because the ELAIS-N1 field uses multiple nights, which have varying ionospheric conditions \citep[see \S~2 in][]{de_jong_into_2024}. A Sigmoid function is fit to the upper 99.9th perecentile of the total flux density to peak brightness ratio as a function of local signal to noise. All sources in the catalogue which lie above this function are treated as resolved, as is any source not fit by a single Gaussian component. 

\section{Methods}
\label{sec:methods}
Brightness temperature, $T_b$, is used to characterise the surface brightness of radio emission. It is not a physical temperature, but rather the temperature of a black body  which would produce the observed spectral radiance (assuming the Rayleigh-Jeans approximation) at long wavelengths. By creating a model which predicts the upper limit of $T_b$ achievable by SF \citep[based on][]{condon_radio_1992}, if the value is \textit{above} that upper limit, the radio emission must be due to AGN activity.  \citetalias{morabito_identifying_2022} showed that $T_b$-based AGN identification can be done at 150 MHz using sub-arcsecond resolution. We make the same conservative assumptions as \citetalias{morabito_identifying_2022}, i.e. we set $T_e=10^4\,$K and the frequency where the optical depth reaches unity, $\nu_0$, to be 3 GHz, for the star formation models. These equate to $T_b\sim10^6\,$K at 144 GHz (equivalent to $T_b\sim10^5\,$K at 1.4 GHz) as the upper limit for SF. The models directly predict flux density per solid angle, which can then be converted to $T_b$ values. Practically, we follow the method in \citetalias{morabito_identifying_2022} which simply uses a threshold in terms of flux density per solid angle rather than converting to $T_b$. We refer the reader to \S~3 of \citetalias{morabito_identifying_2022} for more details. Note that this method yields conservative, positive AGN identifications. 

There are two intermediate cases which may not be caught by this. For the first case take the example of Arp 220, a late-stage merger that has been shown to have a high brightness temperature which is consistent with an origin entirely from densely packed, very luminous radio supernovae \citep{smith_starburst_1998,lonsdale_vlbi_2006}. Sub-arcsecond resolution imaging with the ILT largely supports this \citep{varenius_subarcsecond_2016} but cannot rule out an AGN origin for the outflow in the eastern nucleus. In this extreme case, very high SF in a compact region can mimic AGN-like values of $T_b$. We note that as the models of SF predict \textit{surface brightness}, compactness of SF relative to the resolution element, which in this case is larger than for GHz VLBI, would actually \textit{decrease} the surface brightness as the measured size would be larger than the intrinsic size. To best imitate an AGN, either the SFR has to be extreme and/or the SF has to take place in an area which is approximately the size of the beam.

The second scenario is low-surface brightness AGN emission which mimics SF. This could be due, for example, to restarted radio AGN with a young, prominent core and low surface-brightness extended structures from previous AGN emission \citep[e.g.][]{mingo_revisiting_2019}. Or it could be due to radio emission powered by shocks from AGN winds \citep[e.g.,][]{zakamska_quasar_2014,rankine_placing_2021,petley_connecting_2022}. In either case the radio emission due to SF will be over-estimated, as it is contaminated by low-surface brightness AGN emission.  Based on the number of new $T_b$ identifications in \citetalias{morabito_identifying_2022}, we expect this is $\lesssim$10 percent of cases. 

These two scenarios act in the opposite direction, and may balance each other out in this study. We aim to provide better constraints based on source structure (e.g., compact starbursts) and using a wider range of spatial scales in our ILT imaging \citep[e.g. the 1.\sarc\ 2 resolution imaging in][]{de_jong_into_2024} in future work. 

\subsection{Separation of AGN activity and SF}
\begin{figure}
    \centering
    \includegraphics[width=0.5\textwidth]{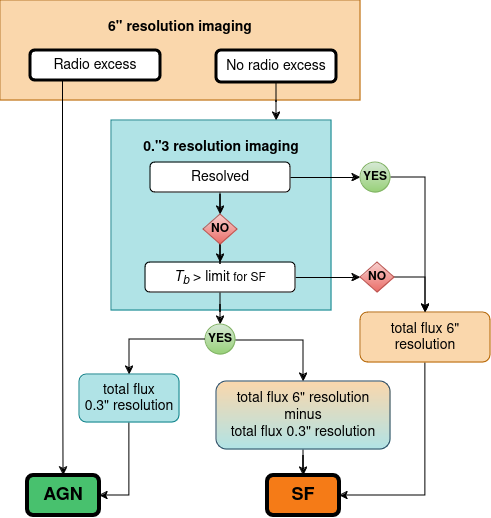}
    \caption{The workflow for determining how to assign flux density from the 6\sarc\ and 0.\sarc\ 3 resolution images to either the AGN or SF category. Only sources with a $T_b$ identified AGN component have their flux density split between the two categories.}
    \label{fig:flowchart}
\end{figure}

Once the AGN activity is identified from the 0.\sarc\ 3 resolution images, it is possible to separate it from SF using the total flux density in the 6\sarc\ \ resolution images. We refer the reader to \S~5 in \citetalias{morabito_identifying_2022} for more details but outline the process here, which is depicted in Fig.~\ref{fig:flowchart}. Sources which are identified as radio excess according to the definition in \citetalias{best_lofar_2023} (and therefore by definition have $>$80 percent of flux density due to AGN) automatically have their total flux density from the 6\sarc\ \ resolution images added to the AGN category. For sources that are not radio excess, we attribute the total flux density from the 6\sarc\ \ resolution images to SF if in the 0.\sarc\ 3 resolution image they are either resolved or have no detection. Sources which do not meet these criteria are checked and if their $T_b$ is not above the limit for SF their 6\sarc\ resolution total flux density is also added to the SF category. The total flux density of any source in the 0.\sarc\ 3 resolution image which is $T_b$-identified as an AGN is assumed to be AGN flux density; this is subtracted from the total flux density measured from the 6\sarc\ \ resolution image, and what remains is assumed to be flux density from SF. In summary, we start with 7,497 total sources. 2,970 have a radio excess, while 4,527 do not. Of those without a radio excess, there are 4,434 unresolved sources for which their $T_b$ is checked. Of those, 1,022 are $T_b$-identified AGN, while 3,412 are not. In the final sample, there are 3,992 sources contributing to the AGN category and 4,527 sources contributing to the SF category. This is more than the total number of sources due to the $T_b$-identified AGN which had a portion of their flux density shifted from the SF to the AGN category.

\subsection{Construction of RLFs}
\begin{figure*}
    \centering
    \includegraphics[width=0.98\textwidth]{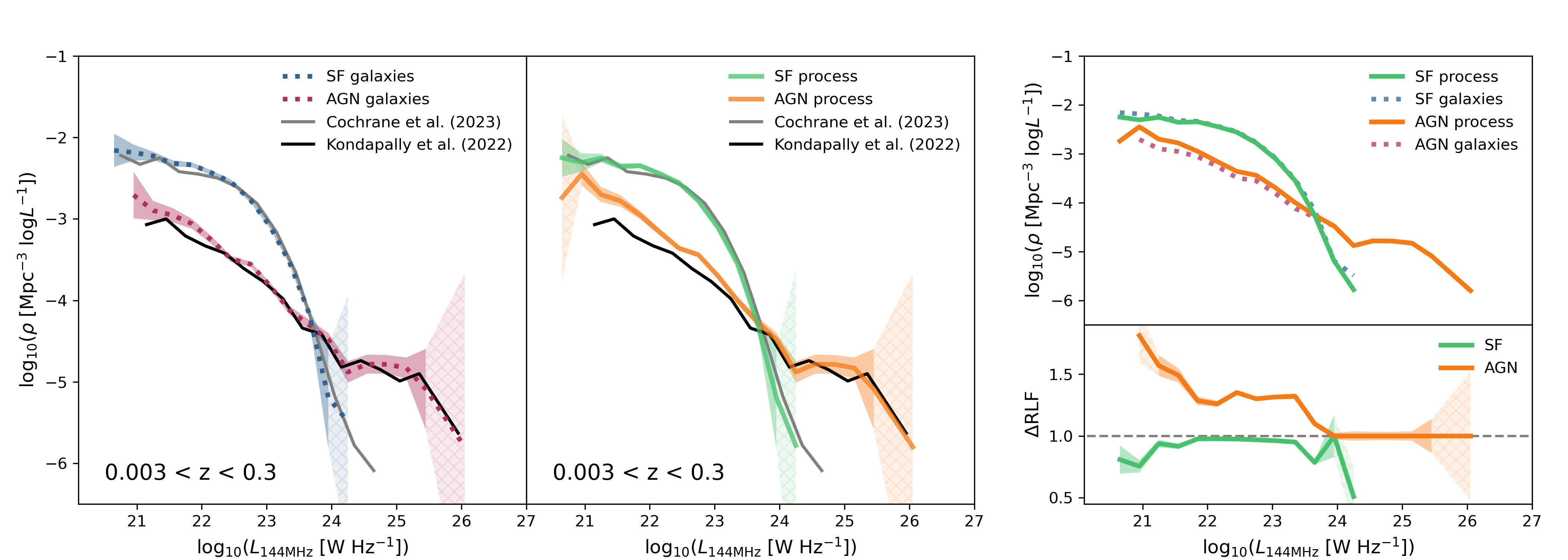}
    \caption{\textit{Left:} The re-calculated galaxy RLFs (dotted lines) for the smaller area considered here, compared with previously published RLFs (solid lines). \textit{Middle:} RLFs calculated by process rather than galaxy. \textit{Right top:} RLFs calculated here by galaxy classifications (dotted lines) and by physical process (solid lines; orange for AGN and green for SF). \textit{Bottom right:} $\Delta$RLF for both AGN and SF. Hatched regions show where the uncertainties are large and values should be treated with caution.}
    \label{fig:rlfs}
    \script{comparison_plot.py}
\end{figure*}

We follow the same method of constructing RLFs as in \cite{kondapally_cosmic_2022} and \cite{cochrane_lofar_2023}, which use the standard 1/$V_{\mathrm{max}}$ method \citep{schmidt_space_1968,condon_14_1989}. The 0.\sarc\ 3 resolution image is smaller than the 6\sarc\ resolution image, so we first reproduce the published RLFs (Fig.~\ref{fig:rlfs}, left panel) using the galaxy classifications from \citetalias{best_lofar_2023} to check that the RLFs in the smaller area are consistent. We use the completeness corrections from the respective papers, with the exception of the photo-$z$ correction used in \cite{cochrane_lofar_2023}. The authors note that this correction is small, and as it is not implemented in \cite{kondapally_cosmic_2022} we chose to keep the method in this paper consistent across both samples. We use the same redshift range, $0.003 < z < 0.3$, to verify that we have good agreement with the published RLFs. Uncertainties are bootstrapped as in \cite{kondapally_cosmic_2022}, with random sampling by replacement to generate 1000 RLF realisations from which the 1$\sigma$ uncertainties are determined. 

Next we use the same 1/$V_{\mathrm{max}}$ method for constructing RLFs by physical process rather than overall galaxy classification, using the flux density of the relevant process, AGN activity or SF, rather than the overall total flux density (where they are not the same). We directly use the completeness corrections from \cite{cochrane_lofar_2023} and \cite{kondapally_cosmic_2022}, based on a galaxy's total flux density. These corrections are flux density dependent, but they are based on 
%we use the total flux density of a galaxy, rather than the flux density of the process, which will be smaller if the radio emission from SF and AGN have been split. This is because the completeness correction is based on 
whether or not a \textit{source} will be detectable given the background rms, not any individual component, although the component flux density can be smaller than the total flux density. 

\section{Results}
\label{sec:results}
The RLFs for the physical processes of SF and AGN are shown in the middle panel of Fig.~\ref{fig:rlfs}, for  $0.003 < z < 0.3$. 
%We present the first RLFs constructed by physical process rather than galaxy classification in Fig~\ref{fig:rlfs}. 
Comparing these with RLFs using galaxy classifications allows us to explore where we are over- or under-predicting the contribution of the physical processes to our statistical studies. To aid this comparison, we define and calculate: 
\begin{equation}
\Delta\textrm{RLF} = \frac{\textrm{RLF(physical process)}}{\textrm{RLF(galaxy classification)}}. \notag
\end{equation}
$\Delta$RLF is calculated once for the SF process/SFG classification, and once for the AGN process/AGN classification. We show $\Delta$RLF in the bottom right hand panel in Fig.~\ref{fig:rlfs}. Above 10$^{24}\,$\wphz\ the ratio for AGN is unity, which is expected as this is where radio excess sources dominate the population. Below that, the ratio for AGN climbs up to $\sim$1.3 and finally almost up to 2 at the lowest radio powers probed. The ratio for SF is always below unity, but it does not deviate as much as the ratio for AGN. This is because the SF population has larger source counts, so the total impact of moving some flux density from the SF to AGN population is less extreme. 

\begin{figure*}
    \centering
    \includegraphics[width=\textwidth]{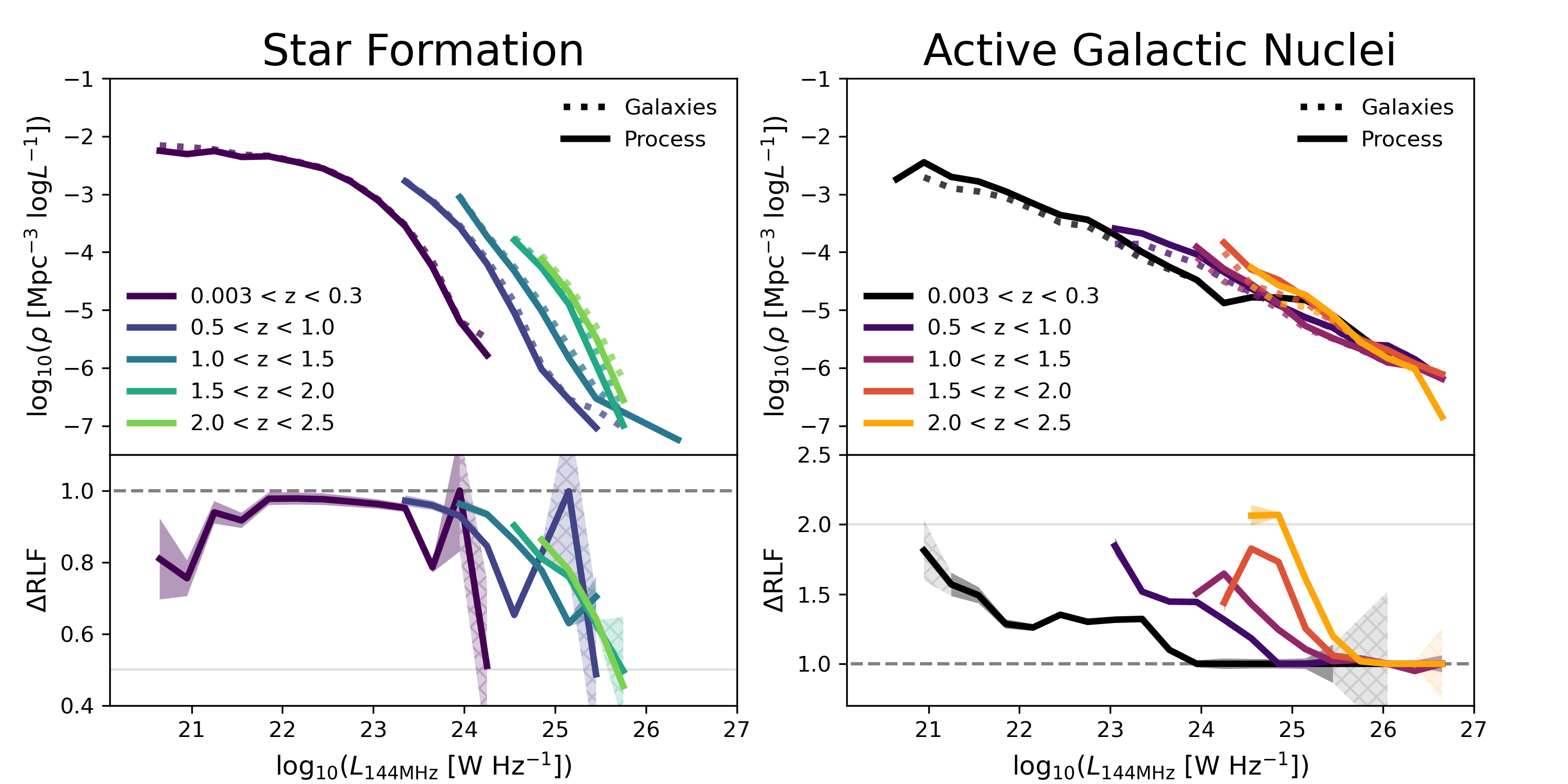}
    \caption{The redshift evolution of the RLFs (top panels) and $\Delta$RLF (bottom panels) for SF (left) and AGN (right). The dashed horizontal line in the bottom panels is unity, with thin solid horizontal lines at 0.5 (left) at 2.0 (right) to guide the eye. To avoid overcrowding, uncertainties are only plotted for $\Delta$RLF, in the same manner as Fig.~\ref{fig:rlfs}. }
    \label{fig:evolution}
    \script{RLF_evolution.py}
\end{figure*}

In Fig.~\ref{fig:evolution} we show the redshift evolution for SF (left) and AGN (right).  We use similar redshift bins to \cite{kondapally_cosmic_2022} and \cite{cochrane_lofar_2023} for consistency. Some bins become unconstrained for the RLFs by process due to the shuffling of sources between the SF and AGN populations, yielding positive y-axis values. We remove these by cutting out bins where there is more than a 4 dex difference between the overall RLF and the process RLF. For the SF RLF we additionally remove points $>-1.8$. We reproduce previous observational results that while the SF population has a strong redshift evolution, the AGN population does not. 

The bottom panels in Fig.~\ref{fig:evolution} show $\Delta$RLF, the same as in the bottom right hand panel of Fig~\ref{fig:rlfs}. For the SF population, $\Delta$RLF is close to unity for $L_{\textrm{144MHz}} < 10^{24}\,$\wphz\ in the lowest redshift bin, and then this begins to drop at higher luminosities. The higher redshift bins show similar behaviour, perhaps reaching unity at increasingly higher values of $L_{\textrm{144MHz}}$, although deeper data is needed to confirm this. 

For the AGN population, $\Delta$RLF has almost the opposite behaviour, converging to unity \textit{above} \llof $\sim10^{24}\,$\wphz\ for $0.003 < z < 0.3$. This is driven by the flux density of radio excess sources, which tend to have higher radio luminosity, being entirely attributed to AGN. The convergence happens at higher luminosity for higher redshift. Below this convergence, $\Delta$RLF increases across all redshift bins. The maximum value for each redshift bin approaches $\sim$2, although in the lowest redshift bin this stabilises at $\sim1.3$ and only climbs higher at the very lowest luminosities (\llof $<10^{22}\,$\wphz ), which may be due to incompleteness. 

To better quantify the integrated effect of $\Delta$RLF, we integrate each RLF separately across the luminosity bins where the uncertainties for both RLFs are well defined, and then take the ratio of the areas. For example, for SF we integrate the RLF for the SF process and the RLF for SFGs, and divide the first by the second. These numbers are always less than unity since the SF process is over-estimated in the SFG population, which has been corrected by moving $T_b$-identified AGN flux density contributions to the AGN process RLF. The integrated values are reported in Tab.~\ref{tab:intvalues}. 
The mean across all redshift bins is 0.90$\pm$0.02 for SF and 1.56$\pm$0.06 for AGN.  
It is evident from the bottom right panel of Fig.~\ref{fig:rlfs} that this `hidden' AGN population extracted from SFGs shows up in the radio AGN population at $L_{\textrm{144MHz}} < 10^{24}\,$\wphz .

\begin{table}
    \centering
    \begin{tabular}{llcc}
 $z_{min}$ & $z_{max}$  & SF & AGN \\ \hline
 0.003 & 0.3 & 0.89$\pm$0.12 & 1.52$\pm$0.36 \\ 
  0.5 & 1.0 & 0.96$\pm$0.06 & 1.53$\pm$0.14 \\ 
  1.0 & 1.5 & 0.95$\pm$0.07 & 1.47$\pm$0.11 \\ 
  1.5 & 2.0 & 0.87$\pm$0.06 & 1.49$\pm$0.36 \\ 
  2.0 & 2.5 & 0.83$\pm$0.06 & 1.80$\pm$0.30 \\ 
     \end{tabular}
    \caption{Integrated $\Delta$RLF, calculated as the ratio of areas under the RLF curve by process to the RLF curve by galaxy classification.}
    \label{tab:intvalues}
\end{table}

\section{Conclusions}
\label{sec:conclusions}
In this letter we presented, for the first time, RLFs constructed by separating radio emission into either SF or AGN processes, for 7,497 sources. We use brightness temperature to identify AGN components in galaxies with no radio excess, and separate the radio emission into AGN and SF. We compare these with classical RLFs constructed using galaxy classifications, and find a hidden AGN population in sources with \llof $\lesssim 10^{24}$\wphz . The mean integrated impact of this is a 1.52$\pm$0.06 increase in the radio luminosity output due to AGN across intermediate redshifts ($0.5 < z < 2.0$), and a doubling in the lowest redshift bin ($0.003 < z < 0.3$). The impact on the SFG population is less severe, amounting to a mean integrated decrease in the RLF to 0.90$\pm$0.02 . The results can have implications for the cosmic SFR density and kinetic luminosity density, which we will explore in a future paper. 

This study is only now possible thanks to LOFAR's exceptional combination of resolution and field of view. The sub-arcsecond resolution is sufficient for brightness temperature identification of AGN, and the field of view means that we have been able to do this for 1,022 sources (out of $>$4,000). We have used conservative limits for AGN identification, which means that there is likely more AGN emission still remaining `uncounted'. Future work to build a forward modelling method, which will also use intermediate resolution images to improve the AGN and SF separation. Splitting radio emission into category by physical process also opens a pathway for direct comparison to cosmological simulations, which we will address in a future paper. 

%\clearpage
%\pagebreak

\section*{Data Availability}
All data used in this paper is either publicly available on \href{https://lofar-surveys.org}{https://lofar-surveys.org} or from relevant citations herein, including catalogues, rms images, and star masks.  This study was carried out using showyourwork (\href{https://github.com/showyourwork/showyourwork}{https://github.com/showyourwork/showyourwork}) which is open source scientific article workflow, first introduced in \cite{luger_mapping_2021}. This leverages continuous integration to
programmatically download the data from \href{https://zenodo.org/}{zenodo.org}, create the figures, and compile the manuscript. The code is available at \href{https://github.com/lmorabit/hidden_AGN}{https://github.com/lmorabit/hidden\_AGN} which is linked to the starting datasets for this work, which are hosted on Zenodo with their own DOIs (10.5281/zenodo.14012620, 10.5281/zenodo.14013377, 14013423, 14013439, 14014490, 14014542, 14014849, and 14014853), ensuring complete reproducibility. 

\section*{Acknowledgements}
LKM recognizes support from UKRI [MR/T042842/1]. 
RK recognizes support from UK STFC via grant [ST/V000594/1].
PNB recognizes support from UK STFC via grants [ST/V000594/1] and [ST/Y000951/1].
BY thanks the University of Edinburgh and Leiden Observatory for support through the Edinburgh–Leiden joint studentship.
JMGHJdJ recognizes support from project CORTEX (NWA.1160.18.316) of research programme NWA-ORC which is (partly) financed by the Dutch Research Council (NWO). 
FS recognizes support from STFC [ST/Y004159/1]. 
MB and IP recognize support from INAF under the Large Grant 2022 funding scheme (project “MeerKAT and LOFAR Team up: a Unique Radio Window on Galaxy/AGN co-Evolution”).
DJBS recognizes support from UK STFC via grants [ST/V000624/1] and  [ST/Y001028/1].
RJvW recognizes support from ERC Starting Grant ClusterWeb 804208. 
This work was supported by SURF Cooperative grant EINF-6218 and EGI-ACE project (Horizon 2020 grant 101017567).
 
The Low Frequency Array was designed and constructed by ASTRON. It has observing, data processing, and data storage facilities in several countries, which are owned by various parties (each with their own funding sources), and collectively operated by the ILT foundation under a joint scientific policy. The ILT resources have benefited from the following recent major funding sources: CNRS-INSU, Observatoire de Paris and Université d'Orléans, France; BMBF, MIWF-NRW, MPG, Germany; Science Foundation Ireland (SFI), Department of Business, Enterprise and Innovation (DBEI), Ireland; NWO, The Netherlands; The Science and Technology Facilities Council, UK; Ministry of Science and Higher Education, Poland; The Istituto Nazionale di Astrofisica (INAF), Italy.

\bibliographystyle{mnras}
\bibliography{references}

\label{lastpage}

\end{document}